\documentclass[overfull,bm]{epl}

\usepackage{amsmath}
\usepackage{amsbsy}
\usepackage{amssymb}

\newcommand{\beq}{\begin{equation}}
\newcommand{\eeq}{\end{equation}}
\newcommand{\bea}{\begin{eqnarray}}
\newcommand{\eea}{\end{eqnarray}}
\newcommand{\nn}{\nonumber}
\newcommand{\openone}{\leavevmode\hbox{\small1\kern-2.8pt\normalsize1}}

\title{Fermionic Casimir effect in case of Andreev reflection}
\shorttitle{Fermionic Casimir effect}

\author{A.  Bulgac\inst{1}
\and
P. Magierski\inst{1,2}
\and
A. Wirzba\inst{3}} 
\shortauthor{A. Bulgac \etal}
\institute{
   \inst{1} Department of Physics, University of Washington - 
            Seattle, WA 98195--1560, USA\\
   \inst{2} Faculty of Physics, Warsaw University of Technology - 
            ul. Koszykowa 75, 00-662 Warsaw, Poland\\
   \inst{3} Institut f\"ur Kernphysik,
            Forschungszentrum J{\"u}lich -
            D-52425 J{\"u}lich, Germany
}

\begin{document}

\maketitle

\begin{abstract}
We describe the Fermionic Casimir effect in the case of two spherical
superfluid scatterers immersed in a normal Fermi system. It is shown
that due to the focusing property of Andreev reflection this new
Casimir-like energy is significantly enhanced when compared to the
case of normal scatterers with specular reflection.
\end{abstract}

\section{Introduction}

If two or more voids (or hard-wall scatterers) are formed in a normal
uniform Fermi system, effective Casimir-like interactions between such
voids appear \cite{abpm,abaw}. For example, two spherical voids of
radius $a$ with distance between their centers $R$ in a Fermi medium
at zero temperature and with Fermi wave vector $k_{\ab{F}}$ would interact
with each other with the effective potential energy 
\beq
E_{\ab{C}}(R) \simeq -\nu_{\ab{deg}} 
 \frac{\hbar^2k_{\ab{F}}^2}{2m} \frac{a^2}{2\pi
R(R-2a)}\,j_1[2k_{\ab{F}}(R-2a)]\,,
 \label{hardsphere}
\eeq
where $\nu_{\ab{deg}}$ is the spin-isospin degeneracy factor.
Here $j_1(x)$ is the spherical Bessel function and $m$ the fermion
mass.  There is an illuminating explanation and easy derivation of
this expression based on semi-classical physics --- periodic orbits
theory in particular.  One can show that the energy density of an
infinite and otherwise homogeneous Fermi system is affected by the
presence of the two spherical voids and that the main contribution to
this energy density change is determined by the ``almost standing
wave'' created between the two scatterers. As  was shown in
refs.~\cite{abpm,abaw}, this argument can be extended, in principle, 
to an arbitrary number of scatterers of arbitrary shapes. Volovik has
arrived at similar conclusions by studying 1-dimensional systems
\cite{volovik} and he refers to this energy as the {\em Mesoscopic Casimir
energy} or {\em force}. 
In his discussion of this new type of Casimir energy,
Volovik stresses that this energy is governed by the ``microscopic
trans-Planckian physics,'' unlike the usual Casimir energy, which is
determined by the long-wave physics.

The Fermionic Casimir energy appears because of the Dirichlet boundary
conditions at the voids, which semi-classically correspond to specular
reflections.  If, however, the boundary separates a normal Fermi system
from a superfluid Fermi system, the specular reflection is replaced
with Andreev reflection \cite{andreev}. A particle 
at the Fermi energy
in the normal part of the Fermi system, impinging on the boundary
between the normal and superfluid system, is retro-reflected as a hole, 
irrespective of the shape of the boundary. This rather simple but
amazing fact has apparently quite unexpected consequences. Let us
imagine now that we have two spherical superfluid grains embedded in a
normal Fermi system. One might then naively expect that there is an
infinite number of {\em stable} (classical) periodic trajectories
between two such objects. This situation has to be contrasted with the
case of specular reflection, when only one {\em unstable} trajectory between
the ``noses'' of the two spheres
exist. The latter trajectory is responsible for the change in the density of
states, which ultimately results in the Fermionic Casimir energy
discussed  in refs.~\cite{abpm,abaw,volovik}. One can now
expect that an infinite number of stable periodic trajectories would
lead to a very large change in the density of states if two superfluid
grains are embedded into a normal Fermi system. From this point of view
the fact that two spherical
superfluid scatterers do not generate a chaotic dynamics is initially
quite a surprise.  Our goal here is to determine the character and
magnitude of the Fermionic Casimir energy in the case of two spherical
superfluid scatterers embedded in a normal Fermi system. The result
we obtain can  easily be generalized to scatterers of arbitrary shapes
and to any number of scatterers  in a leading order
approximation, which we shall define more carefully below. We would
like to mention here that even though  other aspects of the
physics of superfluid grains of various sizes embedded in a normal
Fermi system have been addressed before (see
refs.~\cite{grains,levitov} and references therein), the existence and
nature of the Fermionic Casimir energy was never considered.  Various
other aspects of the quantum chaos in so-called Andreev
billiards/stadiums have been considered by many authors \cite{others},
but spatial configurations of the type we shall study here have
not. Our interest stems partially from the physics of neutron
stars, where such structures are expected to exist
\cite{abpm,seoul}. Moreover, the possible relevance of  Andreev reflection
was mentioned in connection with the appearance of
QCD-color-superfluid ``grains'' in the deconfined quark-gluon phase in
the core of neutron \cite{tachibana} or of quark stars, and also
in connection with the observed cosmological asymmetry between matter
and anti-matter and the origin of dark matter\,\cite{ariel}.

\begin{figure}
\onefigure[width=7.0cm]{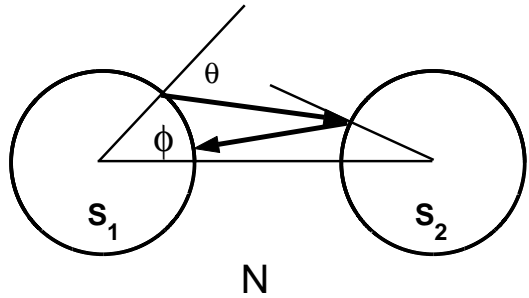}
\caption{
Two superfluid grains  
${\ab{S}}_1$ and ${\ab{S}}_2$ 
immersed in a normal Fermi
gas $\ab{N}$, $\phi$ is the angle at center specifying the position of the
emerging ray on the first disk/sphere and $\theta$ the angle made by
its momentum with the normal.
          }
\label{fig:fig1}
\end{figure}
\section{Two superfluid grains}

To set the stage, let us consider the geometry illustrated in
fig.~\ref{fig:fig1}. In the Andreev approximation any straight line
joining two arbitrary points on the surface of the two superfluid
grains leads to a classical periodic orbit. A particle ($\ab{p}$) leaving the
surface of one grain, hits the second grain somewhere and it is
reflected exactly backward as a hole ($\ab{h}$) and it thus returns to the
initial point on the first grain, where it now transforms, after
another Andreev reflection, into a particle. This is strictly correct
if the particle has exactly the Fermi energy. Then the energy
of both particle and hole states are equal. Otherwise the relation
between the angles of the incoming particle/\-hole and the 
reflected hole/\-particle
resembles Snell's law:
\beq
 k_{\ab{p}} \sin \theta_{\ab{p}} = k_{\ab{h}} \sin \theta_{\ab{h}} \,,
 \quad \frac{k_{\ab{p}}^2}{2} + \frac{k_{\ab{h}}^2}{2} = 2\mu 
  = k_{\ab{F}}^2 \,,
\eeq
where $\mu$ is the Fermi energy and $k_{p}>k_{\ab{F}}$, 
$k_{\ab{h}}<k_{\ab{F}}$ denote
the fermion momentum in a particle state and a hole state,
respectively (we shall use units: $\hbar^2=m=1$).  Note that the
energies of the particle and hole states are symmetric with respect to
the Fermi energy.  Moreover, the incident and reflected rays are on the
same side of the normal to the surface at the point of incidence, as
opposed to the case of specular reflection. Consequently, one can show
that, unlike in the case of normal reflection, the disks/spheres now have a
focusing effect.

Instead of trying to seek various possible periodic trajectories we
simply generated a Poincar\'e surface of section. Only those orbits
are stable for which the separation between the centers of the grains
satisfies the following constraint, derived by requiring that the
eigenvalues of the monodromy matrix have to be inside or 
on the unit circle in the complex plane:
\beq
\frac{R}{2a} \le  \frac{k_{\ab{p}}}{k_{\ab{p}}-k_{\ab{h}}}\, ,
\eeq
where we have assumed that $|k_{p}-k_{h}|/k_{F} \ll 1$.
If this constraint is not satisfied the orbits have a
hyperbolic/unstable character. In the case of equality the eigenvalues
of the monodromy matrix are equal to (minus) one, the orbits are
parabolic/marginally stable in character and elliptic/stable
otherwise.

\begin{figure}
\onefigure[width=7.0cm]{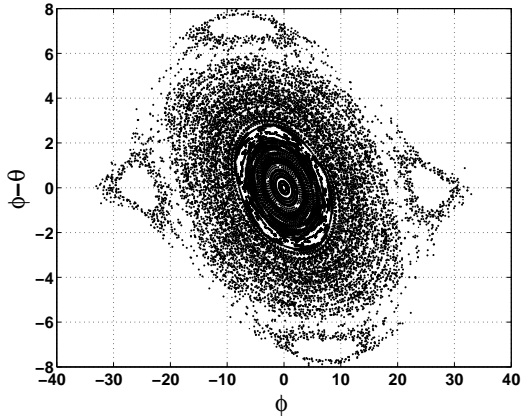}
\caption{ 
The Poincar\'e surface of section for the case of two superfluid
circular/spherical grains embedded in a normal Fermi gas (in two
dimensions), separated by a distance four times larger then their
respective radii ($R$=$6a$). 
The ratio of the particle and hole momenta is here
$k_{\ab{p}}/k_{\ab{h}}$=$1.5$.  
Only stable/\-elliptic trajectories parameterized by the
angle $\phi$ (in degrees), where the particle originates on the
surface of the disk ${\ab{S}}_1$, and by the angle $\phi-\theta$ (in degrees)
between the momentum of the particle and a parallel of the
center-to-center axis are shown. 
The relatively small values of
$\phi-\theta $ show that most trajectories run in a tube around the
line joining the two centers.
}
\label{fig:fig2}
\end{figure}
\section{Exact expression of the density of states}

The structure of the classical phase space briefly described here
would apparently imply a rather complex quantum mechanical
dynamics. As we shall show, however, the most important features of the
density of states in such systems can be understood and described
quite accurately in terms of a simplest periodic orbit and its
repetitions. In order to determine the change in the local density of
states we shall construct explicitly the scattering states and
determine the phase shifts by extending the formalism used in case of
normal systems \cite{abaw,andreas,lloyd} to Bogoliubov--de Gennes
equations,
\beq
\left ( \begin{array}{cc} \tens{T} -\mu & \Delta({\vect{r}}) \\
                          \Delta^*({\vect{r}})& -(\tens{T}-\mu)
\end{array} \right )
\left ( \begin{array}{l} u({\vect{r}}) \\   v({\vect{r}})
\end{array} \right )
= E \left ( \begin{array}{l} u({\vect{r}}) \\   v({\vect{r}})
\end{array} \right )\,,
\eeq
where $\tens{T}$ is the kinetic energy operator, $E$ is the quasi-particle
energy, $u(\vect{r})$ and $v(\vect{r})$ are the particle and hole
spatial components of the quasi-particle wave function and 
$\Delta (\vect{r})$
is the pairing field, assumed to be constant inside the two
scatterers, {\it i.e.} $|\Delta (\vect{r})| = \Delta$, and vanishing
outside. For the sake of simplicity of the analysis we assume that the
mean field  vanishes everywhere and that the two scatterers have
the same radius $a$. Physically it makes sense to discuss many fermion
systems for which $k_{\ab{F}}a \gg 1$. One can distinguish further between
small and large superfluid grains, depending on whether the coherence
length $\xi$ = $k_{\ab{F}}/\pi\Delta$ is smaller or larger than the radius
of the grain $a$. The pairing fields in the two grains need not to
have the same phase, which for the sake of simplicity of the argument 
we consider to be constant over the grain. As in
ref.~\cite{abaw}, we shall use Krein's formula \cite{krein},
generalized to the case of Bogoliubov--de Genes equations. Krein's
formula expresses the change in the density of states through the
phase shifts, and represents a generalization of a much simpler
formula due to Beth and Uhlenbeck \cite{uhlenbeck}. We thus obtain in
the 2-dimensional case (the 3-dimensional case has a very similar structure): 
\bea
 \delta g(E)&=& \frac{\upd\, \delta N(E)}{\upd E}\, , \quad
 \delta N(E)= -\frac{1}{\pi} \ab{Im} \ln \det(\tens{M})\,, \label{eq:exact}\\
 \tens{M} &=& \left (
\begin{array}{cccccc}
...&...&...&... \\
...&\tens{M}_{m\, m'}     & \tens{M}_{m\, m' + 1}&... \\
...&\tens{M}_{m + 1\, m'} & \tens{M}_{m + 1\, m' + 1}&... \\
...&...&...&...
\end{array}
\right ),\nn\\
\tens{M}_{m m'} &=&\left (
\begin{array}{cc}
\openone &    (-1)^{m+1}\tens{A}_{m m'} \\
(-1)^{m+1}\tens{B}_{m m'} &\openone 
\end{array}
\right ) ,\nn\\
\tens{A}_{m m'}&=&\left ( 
\begin{array}{cc}
t_{\ab{pp}}^{m}(a,E,\mu,\Delta)\, H_{m-m'}^{(1)}(k_{\ab{p}}R) &   
            t_{\ab{ph}}^{m}(a,E,\mu,\Delta)\, H_{m-m'}^{(2)}(k_{\ab{h}}R)
\\
t_{\ab{hp}}^{m}(a,E,\mu,\Delta)\, H_{m-m'}^{(1)}(k_{\ab{p}}R) &  
            t_{\ab{hh}}^{m}(a,E,\mu,\Delta)\, H_{m-m'}^{(2)}(k_{\ab{h}}R)
\end{array}
\right ),\nn   \\
\tens{B}_{m m'}&=&\left (
\begin{array}{cc}
t_{\ab{pp}}^{m}(a,E,\mu,\Delta)\, H_{m-m'}^{(1)}(k_{\ab{p}}R)
&{t}_{\ab{ph}}^{m}(a,E,\mu,\Delta) e^{i\phi_\Delta} 
H_{m-m'}^{(2)}(k_{\ab{h}}R)\\
{t}_{\ab{hp}}^{m}(a,E,\mu,\Delta) e^{-i\phi_\Delta}
H_{m-m'}^{(1)}(k_{\ab{p}}R)
&t_{\ab{hh}}^{m}(a,E,\mu,\Delta)\, H_{m-m'}^{(2)}(k_{\ab{h}}R)
\end{array}
\right ),\nn 
\eea
with $  k^2_{\ab{p}}= 2(\mu + E)$,  $ k^2_{\ab{h}}= 2(\mu- E)$ 
and where 
$m,m^\prime = 0,\pm 1,\pm 2, \cdots$ are angular momenta. 
The angle $\phi_\Delta$ is
the relative phase of the pairing fields in the two grains and
$\openone$
is the identity matrix. 
The matrix elements 
$t_{\ab{pp}}^{m}\equiv t_{\ab{pp}}^{m}(a,E,\mu,\Delta), \ldots $ 
are equal to half of the
$T$-matrix elements for the  scattering of a particle/hole of angular 
momentum $m$
off a single spherical superfluid grain (centered at the origin of the
coordinate system),
and the ordinary
Hankel functions $H_{m-m'}^{(1)}(k_{\ab{p}}R), \ldots$ determine the
propagation between the grains. 

\section{Asymptotic approximations}

Even though these formulas are
already suitable  to perform explicit calculations, one can take
advantage of the fact that $k_{\ab{F}} R\gg 1$ and obtain significantly
simpler expressions by retaining only the leading term in an
asymptotic expansion of the corresponding 
Hankel functions. One thus arrives at
the following much more  transparent expressions:
\bea
 T_{\ab{ph}} &=& \Bigl(\sum_{m}t_{\ab{ph}}^{m}\Bigr)^2
\cos \phi_\Delta \equiv
|T_{\ab{ph}}|\exp (i\phi_{\ab{ph}}) \cos \phi_\Delta\,, \nn \\
 T_{\ab{pp}} &=& \Bigl( \sum_{m}t_{\ab{pp}}^{m}(-1)^{m}\Bigr)^2
\equiv |T_{\ab{pp}}|\exp (i\phi_{\ab{pp}})\,, \quad
T_{\ab{hh}}=\Bigl( \sum_{m}t_{\ab{hh}}^{m}(-1)^{m} \Bigr)^2
\equiv |T_{\ab{hh}}|\exp (i\phi_{\ab{hh}})\,, \nn \\
\delta N(E) &\approx&
 \frac{4|T_{\ab{ph}}|\cos\phi_\Delta}{
\pi^2 \sqrt{k_{\ab{h}}k_{\ab{p}}} R}
\sin [(k_{\ab{p}} - k_{\ab{h}})R + \phi_{\ab{ph}}] 
-\frac{2|T_{\ab{pp}}|}{\pi^2 k_{\ab{p}} R} \cos (2 k_{\ab{p}} R 
 + \phi_{\ab{pp}}) \nn\\
&& \mbox{} + \frac{2|T_{\ab{hh}}|}{\pi^2 k_{\ab{h}}R}
\cos (2 k_{\ab{h}} R - 
\phi_{\ab{hh}})\,.
\label{eq:piotr}
\eea
\begin{figure}
\onefigure[width=7.9cm]{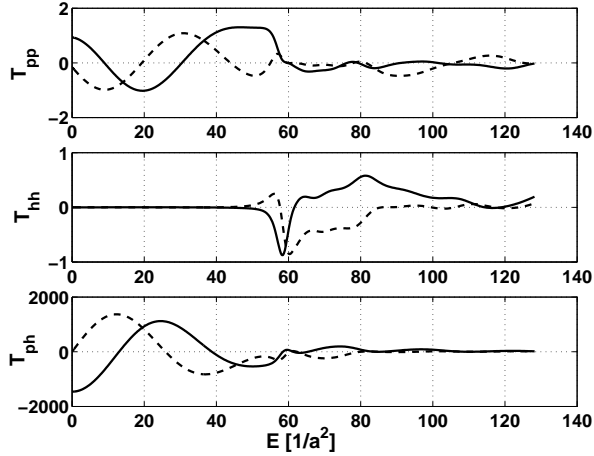}
\caption{ 
The structure functions $T_{\ab{pp}},\; T_{\ab{hh}}$ 
and $ T_{\ab{ph}}$ for the case
of two identical disks with $\mu = 200/a^2$ and $\Delta = 50/a^2$ and
$\phi_\Delta=0$. The solid lines are the real parts of the
corresponding quantities and the dashed lines  the imaginary
parts, respectively. Notice the difference in the magnitudes.
}
\label{fig:fig3}
\end{figure}
The structure functions $ T_{\ab{ph}}$, $T_{\ab{pp}}$ and $T_{\ab{hh}}$ 
characterize
the  properties of the scatterers  and are independent of the
separation $R$ between them. In fig.~\ref{fig:fig3} we show
some typical behavior of these functions with the
quasi-particle energy $E$.
Note the great
difference in size between the normal ($\ab{pp}$, $\ab{hh}$) and Andreev
($\ab{ph}$) structure functions, especially when $E\le \Delta$.
\begin{figure}
\onefigure[width=7.7cm]{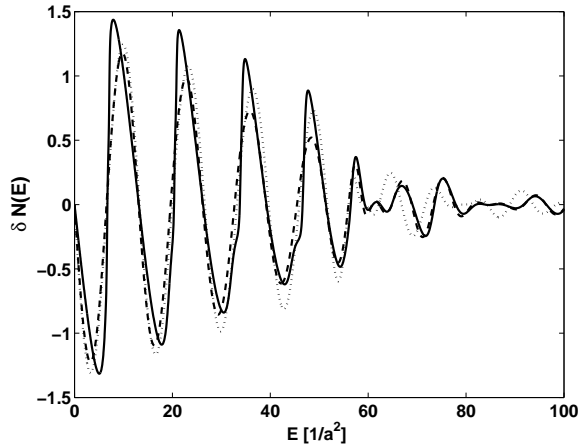}
\caption{ 
The density of states variation $\delta N(E)$ evaluated exactly
(solid line),  in the approximation of eq.~(\ref{eq:piotr})
(dashed line) and under a further asymptotic expansion 
of the cylindrical functions in
the structure functions
(dotted line) for two identical disks with $R$=$6a$,
$\mu$ = 200/$a^2$, $\Delta$=50/$a^2$ and $\phi_\Delta$=0.
}
\label{fig:fig4}
\end{figure}

Even though these structure functions are relatively easy to evaluate
numerically in the case of almost any conceivable size grains, it is
extremely instructive to go one step further and derive an approximate
expression for these structure functions themselves. An analysis of
the structure functions (omitted here) eventually leads us to the following
significantly simpler expression for the change in the density of
states valid for $E \ll |\Delta|$:
\beq
 \delta N(E) = -\ab{Im} \left \{
\left [
\int_{-a \min(k_{\ab{p}},k_{\ab{h}}) }^{a \min(k_{\ab{p}},k_{\ab{h}})} 
\upd\nu\,
\exp \left (\frac{i(k_{\ab{p}}-k_{\ab{h}})\nu^2}
 {2k_{\ab{p}}k_{\ab{h}}a}\right)
\right ]^2  
\frac{e^{i(k_{\ab{p}}-k_{\ab{h}})(R-2a)}}{\pi ^2 R 
 \sqrt{k_{\ab{p}} k_{\ab{h}}}} \right \}
\cos(\phi_\Delta)\,.
\label{eq:andreas}
\eeq
One can easily recognize the argument
of the second exponential as the classical action of the shortest
particle-hole periodic orbit. The hole momentum enters with a minus
sign since it is anti-parallel to the hole velocity. 
The squared integral would become
a product of similar factors in the case of two grains of different
radii. For $E$=0 this overall factor becomes 
$4 m_{\ab{max}}^2 = 4k_{\ab{F}}^2a^2$,
{\it i.e.} the squared number of partial waves undergoing Andreev
reflection.
 
\section{Results and conclusions}
In fig.~\ref{fig:fig4} we compare the change in the
density of states evaluated exactly (see eq.~(\ref{eq:exact})) 
with the simplified expression
of eq.~(\ref{eq:piotr}) and with a further asymptotic approximation of
the structure functions on which eq.~(\ref{eq:andreas}) is based.
It is now obvious that the oscillatory character of the change in the
density of states is fully accounted for by the shortest periodic
orbit, quantized as expected by
\beq
(k_{\ab{p}}- k_{\ab{h}})(R-2a) - 2\arccos\left(E/\Delta\right) =2\pi n\,,
\quad n=0,\pm 1,\pm 2,\cdots\,.
\eeq
The maxima in $\delta g(E)$ = $\upd \delta N(E)/\upd E$ 
correspond exactly to those energies for
which the above condition is fulfilled.  At each Andreev reflection
the reflected wave acquires an additional phase $\arccos(E/\Delta)$,
leading to the  overall minus sign in eq.~(\ref{eq:andreas}), valid for the
case $E\ll \Delta$.

By using the exact expression (\ref{eq:exact}) or the simplifications
(\ref{eq:piotr}) or (\ref{eq:andreas}),  one can now 
easily evaluate the interaction energy of spherical  
superfluid grains in the Fermi sea, 
\beq
E_{\ab{SC}}(R) = \nu_{\ab{deg}} \int_0^{\infty}  \upd E\, 
                   E \,\frac {\upd\, \delta N(E)}{\upd E}
               =-\nu_{\ab{deg}} \int_0^{\infty}  \upd E\, \delta N(E) \, ,
\eeq
as function of their separation $R$.
This Casimir energy is completely  dominated  by the
states with $E\le \Delta$.
We thus obtain, by numerical integration, for 
two spherical three-dimensional grains 
\beq
E_{\ab{SC}}(R) \approx \nu_{\ab{deg}}\frac{\hbar^2 k_{\ab{F}}^2}{2m}\, 
\frac{k_{\ab{F}} a^4 \cos\phi_{\Delta} }{2 \pi R^2(R-2a)} X \,, 
\eeq
where $X={\cal{O}}(1)$ and positive and 
very weakly dependent on separation. By comparing with eq.~(\ref{hardsphere}),
one can show that
$E_{\ab{SC}}(R)\gg |E_{\ab{C}}(R)|$ for the 
typical case  $k_{\ab{F}} a\gg 1$.
This is easy to accept now:
in the case of two superfluid grains, embedded in a normal Fermi system, there are
many  ``bound states'' (see fig.~\ref{fig:fig2}) due to the
appearance of a large number of periodic orbits, whereas 
there exists only a single unstable periodic orbit between  two hard 
spheres  \cite{abpm,abaw}. 

In summary, we have constructed both {\em exact} and  {\em simplified}
expressions for the
calculation of the density of states and  the  Fermionic 
Casimir energy in the novel case of
two spherical superfluid scatterers immersed in a normal Fermi system.
We have furthermore shown and explained 
that these quantities are significantly enhanced
in comparison to the normal case where the obstacles are 
scatterers with hard-wall reflection conditions.
Since the change in the density of states is dominated by the Andreev
reflection, which is almost exactly backward, the case of many grains
reduces simply to the sum over all unobstructed pairs of grains. 
In particular, 
one can expect that the energetics in portions of the
neutron stars would be greatly affected when the Fermionic Casimir
energy is taken into account --- more so than was the case when only
normal systems were considered \cite{abpm,seoul}. 
Generalizations of
the present results to more complicated geometries appear to be 
straightforward.

\acknowledgements
This work was supported in part by DOE grant DE-FG02-97ER41014, by the 
Polish Committee for Scientific Research (KBN) under Contract No.~1~P03B~059~27
and by the Forschungszentrum J\"ulich under Contract No.~41445400 (COSY-067).
We would like to thank one of the referees for valuable suggestion.

\end{document}